\documentclass[twocolumn,showpacs,preprintnumbers,nofootinbib]{revtex4-1}

\usepackage{graphicx}  
\usepackage{dcolumn}
\usepackage{bm}
\usepackage{amsmath,amssymb,amsfonts}
\usepackage{latexsym}
\usepackage{color}

\def\nn    {\nonumber}

\begin{document}

\title{\boldmath
Prospect and implications of $cg\to bH^+\to b A W^+$ production at the LHC}

\author{Wei-Shu Hou$^1$, Tanmoy Modak$^2$}
\affiliation{
$^1$Department of Physics, National Taiwan University, Taipei 10617, Taiwan\\
$^2$Institut f{\"u}r Theoretische Physik, Universit{\"a}t Heidelberg, 69120 Heidelberg, Germany}

\bigskip

\begin{abstract}
We study the prospect for discovering the $cg\to bH^+\to b A W^+$ process at the LHC.
Induced by the top-flavor changing neutral Higgs coupling $\rho_{tc}$, 
the process may emerge if $m_{H^+} > m_A + m_{W^+}$,
 where $H^+$ and $A$ are charged and $CP$-odd Higgs bosons 
in the general two Higgs Doublet Model (g2HDM). 
We show that the $cg\to bH^+\to b A W^+$ process can be
discovered at LHC Run 3, while the full Run 2 data at hand can 
constrain the parameter space significantly by searching for the same-sign dilepton final state. 
The process has unique implications on the hint of $gg\to A \to t \bar t$ 
excess at $m_A\approx 400$ GeV reported by CMS. 
When combined with other existing constraints, 
the $cg\to bH^+\to b A W^+$ process can essentially rule out 
the g2HDM explanation of such an excess.
\end{abstract}

\maketitle

\section{Introduction}

The discovery of the 125 GeV Higgs boson $h$~\cite{h125_discovery} 
at the Large Hadron Collider (LHC) and subsequent measurements of its 
couplings~\cite{higsscoup} confirm that the Standard Model (SM) 
is the correct effective theory at the electroweak scale. 
While there is no compelling experimental evidence so far for new physics 
(NP) beyond SM, additional Higgs bosons may well exist in Nature. 
Most ultraviolet (UV) models 
have extensions of the Higgs sector, while their effective descriptions 
at sub-TeV scale should resemble the SM. 
The two Higgs doublet model (2HDM)~\cite{2hdmreview}, 
with two scalar doublets $\Phi$ and $\Phi'$, 
is one of the simplest renormalizable extensions of the SM. 
While the extra scalars could be at 
the so-called decoupling limit~\cite{Gunion:2002zf} and heavy, 
more interesting is when they are sub-TeV in mass, 
with the $h$ boson couplings to fermions and gauge bosons SM-like 
as observed~\cite{Biekotter:2018rhp,Ellis:2018gqa,Almeida:2018cld}.

Our context would be the general two Higgs doublet model (g2HDM).
Unlike the popular 2HDM-II (which automatically arises with supersymmetry),
in the absence of any discrete symmetry, both the $\Phi$ and $\Phi'$ doublets 
couple to $F = u$ and $d$-type quarks (as well as charged leptons).
After diagonalization of the fermion mass matrices,
two separate Yukawa matrices,
 $\lambda_{ij}^F =  \delta_{ij} {\sqrt{2}m_i^F}/{v}$
 (with $v \simeq 246$ GeV) and $\rho_{ij}^F$, emerge.
The $\lambda$ matrices are real and diagonal as in SM,
but the $\rho$ matrices are in general nondiagonal and complex.
It has been shown that complex $\rho_{tt}$~\cite{Fuyuto:2017ewj} and $\rho_{bb}$~\cite{Modak:2018csw} can each account for 
the observed matter-antimatter asymmetry via electroweak baryogenesis (EWBG).
Our focus of interest, however, would be the flavor changing neutral Higgs 
(FCNH) coupling $\rho_{tc}$, which is found~\cite{Fuyuto:2017ewj} to be 
also capable of driving EWBG~\cite{Fuyuto:2017ewj}
when ${\cal O}(1)$ and with near-maximal phase.

Despite the attraction of EWBG, and the fact that we have
the least knowledge about extra top Yukawa couplings, 
it has been raised long ago~\cite{Glashow:1976nt} the preferred absence of
flavor changing neutral couplings (FCNC), such as $\rho_{tc}$.
It is customary, therefore, to invoke a $Z_2$ symmetry to enforce 
the Natural Flavor Conservation (NFC) condition~\cite{Glashow:1976nt}, 
that there be only one Yukawa matrix even in 2HDM context.
Caution was first raised by Cheng and Sher~\cite{Cheng:1987rs} 
that the NFC condition may be overkill,
and e.g. $\rho_{ij} \propto \sqrt{m_i m_j}/v$, which reflects
the quark mass and mixing hierarchies, could help alleviate
the concerns of Ref.~\cite{Glashow:1976nt}.
As the pattern implies $\rho_{tc}$ would be the largest FCNC,
thereby anticipating~\cite{Hou:1991un} future $t \to ch$ or $h \to t\bar c$ search,
it was asserted that indeed the mass-mixing hierarchies
illustrate Nature's ``design'', while one does not have to
adhere to the Cheng-Sher ansatz strictly.

More recently, with the SM-like $h(125)$ lighter than the top
 --- whereby ATLAS and CMS immediately started $t \to ch$
     (and also $h \to \tau\mu$) search~\cite{PDG}
 --- one notes~\cite{Chen:2013qta} that the $tch$ coupling should be 
modulated by $\cos\gamma \equiv c_\gamma$, the $h$-$H$ mixing angle 
between the two $CP$-even Higgs bosons of 2HDM.
With subsequent emergence of the ``alignment'' phenomenon~\cite{higsscoup},
that $h$ resembles the SM Higgs boson so well,
a further non-flavor mechanism was added to Nature's ``design''
for hiding the effects of tree level FCNC's arising from the Higgs sector:
small $c_\gamma$.
Indeed, one may not need the {\it ad hoc} NFC condition.

The FCNH coupling $\rho_{tc}$ can be discovered at the LHC 
via the $cg\to t A/tH \to t t \bar c$ process, i.e. the same-sign top 
signature~\cite{Kohda:2017fkn,Hou:2018zmg}
 (see also Refs.~\cite{Hou:1997pm,sstother}). 
With both top quarks decaying semileptonically, 
the $cg\to t A/tH \to t t \bar c$ process provides a clean discovery mode for $\rho_{tc}$, 
even for $c_\gamma = 0$. 
For moderate $c_\gamma$ values, one may also have 
the $cg\to b H^+\to b h W^+$ process, which provides another sensitive 
probe for $\rho_{tc}$~\cite{Hou:2020tnc} (see also Ref.~\cite{Gori:2017tvg}).

\begin{figure}[b!]
\center
\includegraphics[width=.37 \textwidth]{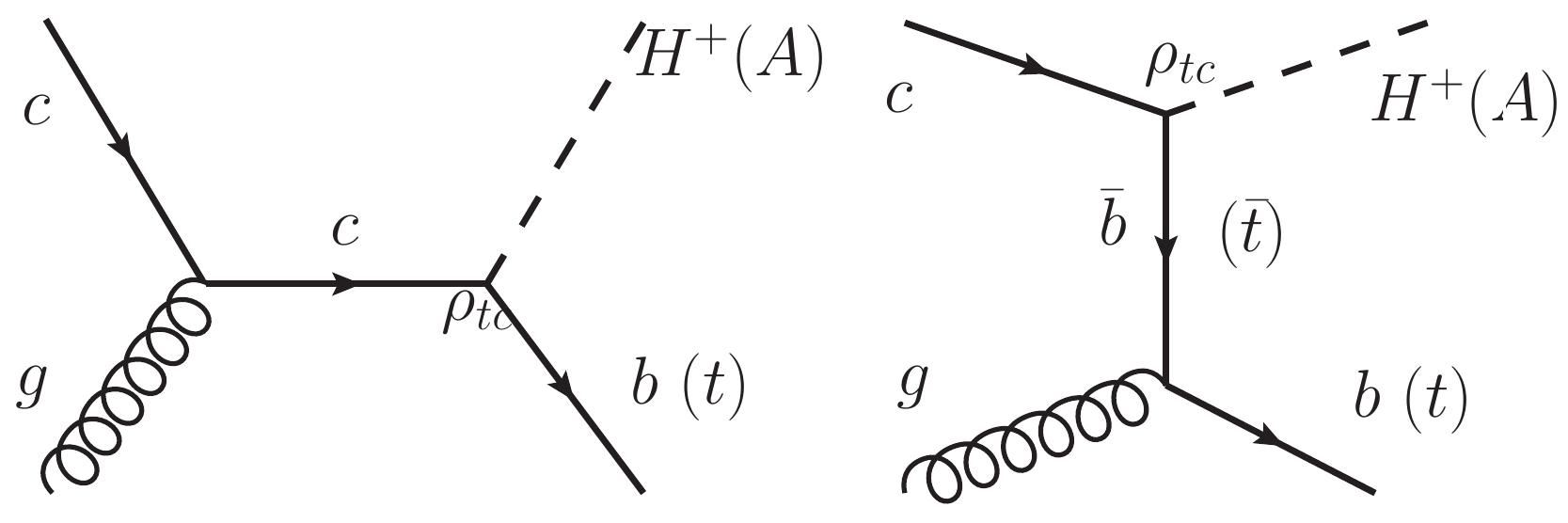}
\caption{
Representative diagrams for $\rho_{tc}$ induced $cg \to b H^+$ ($t A$) process.} 
\label{cgbhp}
\end{figure}

In this article we study the prospect of probing $\rho_{tc}$ 
via the novel $cg\to b H^+\to b AW^+$ process (conjugate process implied).
The production of $cg\to b H^+$~\cite{Iguro:2017ysu,Ghosh:2019exx} is 
initiated by the $\rho_{tc}$ coupling (see Fig.~\ref{cgbhp}), 
while $H^+ \to AW^+$ decay can occur for $m_{H^+} > m_A + m_{W^+}$. 
Like the same-sign top signature, the process also does not 
depend on the mixing angle $c_\gamma$. 
We study the $cg\to b H^+\to bAW^+$ process
followed by $\rho_{tc}$-induced $A\to t \bar c$ decay at 14 TeV LHC.  
With semileptonic decay of $t$ and leptonic decay of $W^+$,
the $cg\to b H^+\to bAW^+$ process could provide 
complementary probe for $\rho_{tc}$, and therefore 
shed light on the $\rho_{tc}$-driven EWBG mechanism.

We analyze also the impact of the $cg\to b H^+\to bAW^+$ process 
on the CMS hint for $gg\to A \to t \bar t$ excess~\cite{Sirunyan:2019wph}
at $m_{t\bar t}\,\sim\,400$\,GeV. 
The ``excess'' can be explained in g2HDM~\cite{Hou:2019gpn} 
if $\rho_{tt}\approx 1.1$ and $\rho_{tc}\approx 0.9$ 
with $m_H\, (m_{H^+})\gtrsim 500\, (530)$ GeV. 
But this parameter range would induce also the $cg\to b H^+\to bAW^+$ process, 
which we show that it would contribute abundantly to some 
control region of an existing CMS search~\cite{Sirunyan:2019wxt}, 
hence can essentially exclude the g2HDM explanation of such an excess.

This paper is organized as follows. 
In Sec.~\ref{sec:param} we discuss the framework and available parameter space. 
In Sec.~\ref{coll} we study the prospect for the $cg\to b H^+\to bAW^+$ process at the LHC. 
Sec.~\ref{cmsexces} is dedicated to the impact of $cg\to b H^+\to bAW^+$ 
on interpreting the CMS ``excess'' in $gg\to A\to t \bar t$. 
We conclude with some discussions in Sec.~\ref{disc}.

\section{Framework and parameter Space}\label{sec:param}

\subsection{\boldmath Relevant interactions}

The most general $CP$-conserving two Higgs doublet potential 
can be written as~\cite{Davidson:2005cw,Hou:2017hiw}
\begin{align}
 & V(\Phi,\Phi') = \mu_{11}^2|\Phi|^2 + \mu_{22}^2|\Phi'|^2
            - (\mu_{12}^2\Phi^\dagger\Phi' + h.c.) \nn \\
 & \quad + \frac{\eta_1}{2}|\Phi|^4 + \frac{\eta_2}{2}|\Phi'|^4
           + \eta_3|\Phi|^2|\Phi'|^2  + \eta_4 |\Phi^\dagger\Phi'|^2 \nn \\
 & + \left[\frac{\eta_5}{2}(\Phi^\dagger\Phi')^2
     + \left(\eta_6 |\Phi|^2 + \eta_7|\Phi'|^2\right) \Phi^\dagger\Phi' + h.c.\right],
\label{pot}
\end{align}
in the Higgs basis,
where the $\eta_i$s are the quartic couplings
and we follow the notation of Ref.~\cite{Hou:2017hiw}. 
The vacuum expectation value $v$ arises from the doublet $\Phi$ 
via the minimization condition $\mu_{11}^2=-\frac{1}{2}\eta_1 v^2$, 
while $\left\langle \Phi'\right\rangle =0$ (hence $\mu_{22}^2 > 0$) 
and the second minimization condition is $\mu_{12}^2 = \frac{1}{2}\eta_6 v^2$. 
The mixing angle $\gamma$ diagonalizes the mass-squared matrix for
$h$, $H$, and satisfies~\cite{Davidson:2005cw,Hou:2017hiw}
\begin{align}
 c_\gamma^2 = \frac{\eta_1 v^2 - m_h^2}{m_H^2-m_h^2},~\quad \quad \sin{2\gamma} = \frac{2\eta_6 v^2}{m_H^2-m_h^2}.
\end{align}
In the alignment limit of $c_\gamma \to 0$, $h$ approaches the SM Higgs boson.
The scalar masses can be expressed in terms of the parameters in Eq.~(\ref{pot}),
\begin{align}
 &m_{h,H}^2 = \frac{1}{2}\bigg[m_A^2 + (\eta_1 + \eta_5) v^2\nn\\
 &\quad\quad \quad\quad \mp \sqrt{\left(m_A^2+ (\eta_5 - \eta_1) v^2\right)^2 + 4 \eta_6^2 v^4}\bigg],\\
 &m_{A}^2 = \frac{1}{2}(\eta_3 + \eta_4 - \eta_5) v^2+ \mu_{22}^2,\\
 &m_{H^+}^2 = \frac{1}{2}\eta_3 v^2+ \mu_{22}^2.
\end{align}

The scalar bosons $h$, $H$, $A$ and $H^+$ in g2HDM couple to 
fermions by~\cite{Davidson:2005cw,Hou:2019mve}
\begin{align}
\mathcal{L} = 
-&\frac{1}{\sqrt{2}} \sum_{F = U, D, L}
 \bar F_{i} \bigg[\big(-\lambda^F_{ij} s_\gamma + \rho^F_{ij} c_\gamma\big) h \nn\\
& \;\ +\big(\lambda^F_{ij} c_\gamma + \rho^F_{ij} s_\gamma\big)H -i ~{\rm sgn}(Q_F) \rho^F_{ij} A\bigg]  P_R F_{j}\nn\\
 &-\bar{U}_i\big[(V\rho^D)_{ij} P_R-(\rho^{U\dagger}V)_{ij} P_L\big]D_j H^+ \nn\\
 &- \bar{\nu}_i\rho^L_{ij} P_R L_j H^+ +{\rm H.c.},\label{eff}
\end{align}
where $P_{L,R}\equiv (1\mp\gamma_5)/2$, $i,j = 1, 2, 3$ are generation indices, 
$V$ is the Cabibbo-Kobayashi-Maskawa (CKM) matrix, 
whereas in flavor space, the $U$, $D$ and $L$ matrices are defined as 
$U=(u,c,t)$, $D = (d,s,b)$, $L=(e,\mu,\tau)$ and $\nu=(\nu_e,\nu_\mu,\nu_\tau)$. 
The matrices
$\lambda^F_{ij}\; (\equiv \delta_{ij}\sqrt{2}m_i^F/v)$ are diagonal and real, while
$\rho^F_{ij}$ are in general complex and nondiagonal. In what follows we shall
drop the superscript $F$ for simplicity.

We are interested in $cg\to b H^+ \to bAW^+$, where $\rho_{tc}$ 
induces $cg\to b H^+$ production (Fig.~\ref{cgbhp}), 
as one can see from Eq.~\eqref{eff}. 
Unlike $Z_2$ symmetric cases such as 2HDM type-II, 
intriguingly the production in g2HDM is CKM-enhanced, 
$V_{tb} \rho_{tc}$~\cite{Ghosh:2019exx}. 
There exist several direct and indirect constraints on $\rho_{tc}$ 
which we shall return shortly. 
The decay $H^+\to A W^+$ on the other hand arises through
%
\begin{align}
 -\frac{g_2}{2} \left(A \partial^\mu H^+ -  H^+ \partial^\mu A   \right) W^-_\mu + \mbox{H.c.},
\end{align}
where $g_2$ is $SU(2)$ gauge coupling. Note that 
the $cg\to b H^+ \to bAW^+$ process is independent of the mixing angle $c_\gamma$,  
while we consider $A \to t\bar c \to bW^+\bar c$ final state, 
with both $W^+$ bosons decaying leptonically.

\subsection{Constraints on parameter space}
\label{sec:Constraints}

There exist several direct and indirect constraints on $\rho_{tc}$.
For $c_\gamma \neq 0$, $\rho_{tc}$ is constrained by 
$t\to c h$ search, i.e. the bounds on $\mathcal{B}(t\to c h)$. 
We take
\begin{align}
  \mathcal{B}(t\to c h) \approx \frac{c_\gamma^2 |\rho_{tc}|^2}{7.66 + c_\gamma^2 |\rho_{tc}|^2}\label{ttoch},
\end{align}  
where we approximate the total width of $t$ quark as the sum of 
$t \to bW^+$ and $t \to ch$ partial widths.
Both ATLAS and CMS have searched for the $t\to ch$ decay 
and set 95\% CL upper limits on the the branching fraction.
The latest ATLAS limit is $\mathcal{B}(t\to c h) < 1.1\times 10^{-3}$, 
based on 36.1 fb$^{-1}$ data~\cite{Aaboud:2018oqm} at 13 TeV,
while the CMS limit of $\mathcal{B}(t\to c h) < 4.7 \times 10^{-3}$, 
based on similar dataset~\cite{Sirunyan:2017uae}, is weaker. 
We find that $|\rho_{tc}| \gtrsim 0.3$ is 
excluded at $95\%$ CL for $c_\gamma \sim 0.3$. 
The limit weakens for smaller $c_\gamma$ and vanishes in the alignment limit.

There are also constraints from flavor physics. 
For example, $\rho_{tc}$ enters through loops with charm quarks
and a charged Higgs into $B_{s}$-$\overline{B}_{s}$ mixing and
$\mathcal{B}(B\to X_s\gamma)$~\cite{Altunkaynak:2015twa}. 
Reinterpreting the limits from Ref.~\cite{Crivellin:2013wna}, we find that 
$|\rho_{tc}|\gtrsim 0.9~(1.2)$ is excluded for $m_{H^+}= 300~(500)$ GeV. 
For the ballpark $m_{H^+}$ values we shall consider, 
the flavor constraint is rather weak. 

The most stringent limit on $\rho_{tc}$ turns out to be
the CMS search for four-top production~\cite{Sirunyan:2019wxt}, 
and comes from the Control Region for $t\bar tW$ background, called CRW. 
With the signature of a same-sign dilepton pair, two $b$-tagged jets and $E_T^{\rm{miss}}$, 
the $cg\to b H^+ \to bAW^+ \to bt\bar cW^+ \to bb\bar c W^+W^+$ process 
would contribute to CRW abundantly.
This is similar to the four-top constraint placed on the $cg\to t A/tH \to t t \bar c$ processes~\cite{Hou:2018zmg,Hou:2019gpn}, 
which have identical final state topologies if 
both of the same-sign top quarks decay semileptonically. 
We shall therefore give a detailed collider study in Sec.~\ref{coll}.

At this point we also remark that the process $cg\to b H^+ \to bAW^+$ can also 
be induced by $\rho_{ct}$ for which a similar search strategy can be adopted. 
In what follows we set all $\rho_{ij} =0$ except $\rho_{tc}$ for simplicity, 
with the impact of other $\rho_{ij}$'s discussed later in the paper. 
Furthermore, as the $cg\to b H^+ \to bAW^+$ does not depend on $c_\gamma$, 
we simply assume the alignment limit and set $c_\gamma =0$ throughout the paper.


The $cg\to b H^+ \to bAW^+$ process requires $m_{H^+} > m_A + m_{W^+}$. 
Before exploring this mass spectrum,  
one needs to ensure the dynamical parameters in Eq.~\eqref{pot}
satisfy perturbativity, tree-level unitarity and vacuum stability conditions, 
for which we use the public tool 2HDMC~\cite{Eriksson:2009ws}. 
We express the quartic couplings $\eta_1$, $\eta_{3{\rm -}6}$ in terms of 
$m_h^2$, $m_H^2$, $m_{H^+}^2$, $m_A$, $\mu_{22}^2$, $\gamma$, 
and $v$~\cite{Davidson:2005cw}, i.e.
\begin{align}
& \eta_1 = \frac{m_h^2 s_\gamma^2 + m_H^2 c_\gamma^2}{v^2},\\
& \eta_3 =  \frac{2(m_{H^+}^2 - \mu_{22}^2)}{v^2},\\
& \eta_4 = \frac{m_h^2 c_\gamma^2 + m_H^2 s_\gamma^2 -2 m_{H^+}^2+m_A^2}{v^2},\\
& \eta_5 =  \frac{m_H^2 s_\gamma^2 + m_h^2 c_\gamma^2 - m_A^2}{v^2},\\
& \eta_6 =  \frac{(m_h^2 - m_H^2)(-s_\gamma)c_\gamma}{v^2}.
\end{align}
The quartic couplings $\eta_2$ and $\eta_7$ do not enter scalar masses.
Imposing $m_{H^+} > m_A + m_{W^+}$, 
we randomly generate the phenomenological parameters 
$\gamma$, $m_A$, $m_H$, $m_{H^+}$, $\mu_{22}$, $\eta_2$, $\eta_7$ 
in the following ranges:
$\mu_{22} \in [0, 1000]$\,GeV,
$m_{H^+} \in [300, 600]$\,GeV,
$m_A \in [200, 600-m_{W}]$\,GeV,
$m_H = m_{H^+}$, 
$\eta_2 \in [0, 5]$, $ \eta_7 \in [-5, 5]$, 
with $m_h = 125$\,GeV and $c_\gamma = 0$ held fixed.

\begin{figure}[t]
\centering
\includegraphics[width=.45 \textwidth]{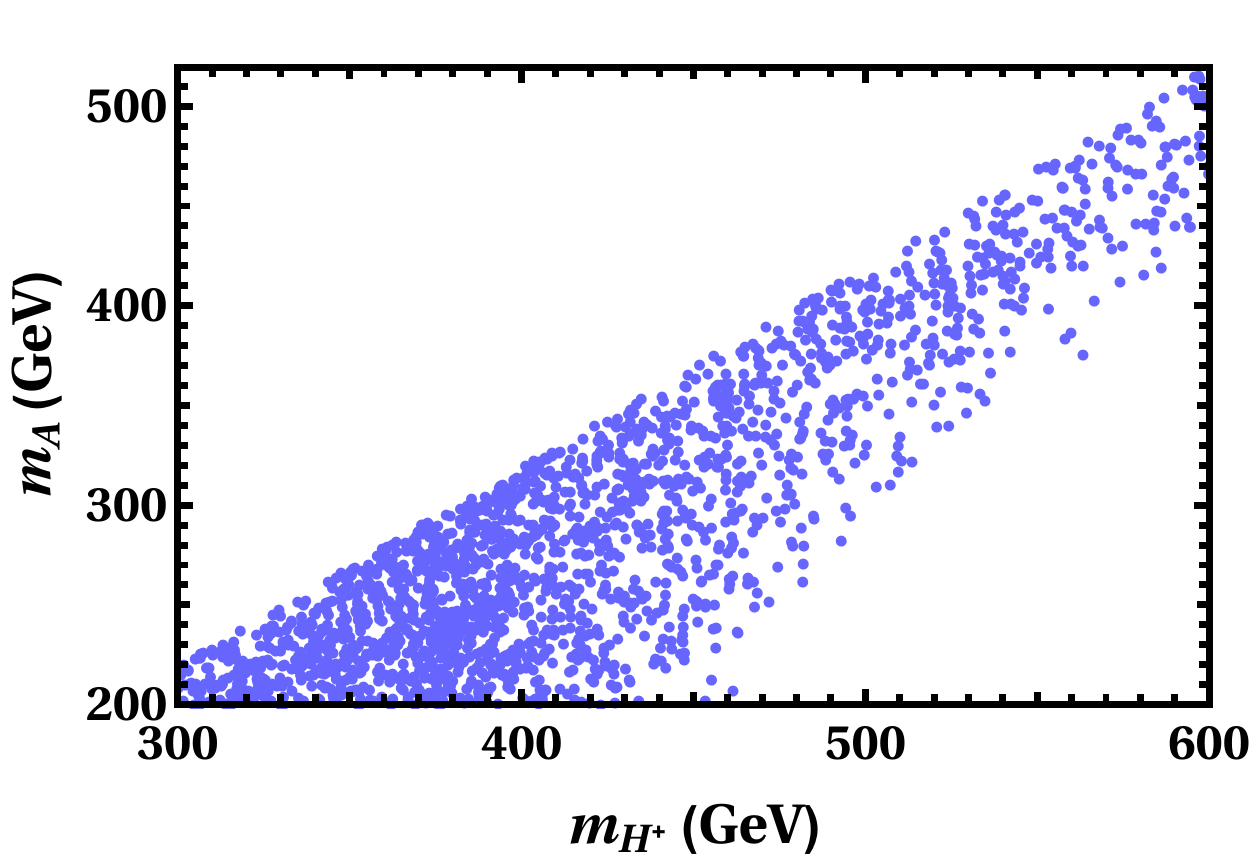}
\caption{
Scan points satisfying $m_{H^+} > m_{A} + M_{W^+}$ and
consistency conditions in the $m_{H^+}$--$m_A$ plane. 
See text for details.}
\label{scan}
\end{figure}

The randomly generated parameters are fed into 
2HDMC~\cite{Eriksson:2009ws} for scanning. 
2HDMC uses $\Lambda_{1-7}$ and $m_{H^+}$ as input parameters 
in the Higgs basis with $v \simeq 246$ GeV. 
We identify $\eta_{1-7}$ as $\Lambda_{1-7}$ and take $-\pi/2\leq \gamma \leq \pi/2$. 
For positivity of the Higgs potential, Eq.~\eqref{pot}, one requires $\eta_2>0$, 
along with other more involved conditions implemented in 2HDMC. 
We further conservatively demand $|\eta_i| \leq 5$.
These scan points are plotted in the $m_{H^+}$--$m_A$ plane in Fig.~\ref{scan}.

One also needs to consider constraints from electroweak precision~\cite{Peskin:1991sw} observables, which further restricts the hierarchical structures 
between the scalar masses $m_H$, $m_A$ and $m_{H^+}$~\cite{Froggatt:1991qw,Haber:2015pua}, and therefore $\eta_i$s. 
For sake of simplicity, we have taken $m_H = m_{H^+}$, 
which corresponds to twisted custodial symmetry~\cite{Gerard:2007kn}.
In general, for non-degenerate $m_H$ and $m_{H^+}$, 
the randomly generated parameters that passed 
unitarity, perturbativity and positivity conditions from 2HDMC, 
can easily be tested for the oblique parameter constraints~\cite{Baak:2014ora} 
also in 2HDMC.\footnote{
Further details on parameter counting and the scanning procedure 
can be found in Refs.~\cite{Hou:2019qqi,Modak:2019nzl,Hou:2019mve,Modak:2020uyq}.
}

It is clear from Fig.~\ref{scan} that there exist a significant 
range of scan points that can facilitate $H^+ \to A W^+$ decay. 
We choose two benchmark points (BPs) from Fig.~\ref{scan} for illustration, 
and list the parameter values in Table~\ref{benchm}.

\begin{table*}[t!]
\centering
\begin{tabular}{c |c| c| c| c | c | c| c | c |c| c |c  c c| cc}
\hline
BP  & $\eta_1$ &  $\eta_2$   &  $\eta_3$   & $\eta_4$  & $\eta_5$ & $\eta_6$ 
 & $\eta_7$ & $m_{H^+}$  & $m_A$ & $m_H$ &  $\frac{\mu_{22}^2}{v^2}$\\
 &&&&&&&& (GeV) & (GeV) & (GeV)&\\ 
\hline
&&&&&&&&&&&\\
$a$  & 0.258  & 2.79  & 2.279  & $-1.342$ & 1.342 & 0
 & $-1.671$ & 354  & 210 & 354 & 0.93\\
$b$  & 0.258  & 2.31  & 3.966  & $-2.061$ & 2.061 & 0
 & $-1.171$ & 531  & 396 & 531 & 2.67\\
\hline
\hline
\end{tabular}
\caption{
Parameter values for the two benchmark points chosen from the scan points in Fig.~\ref{scan}.
}
\label{benchm}
\end{table*}

\section{Prospect at the LHC}\label{coll}

We now discuss the constraint from the CRW region 
of the CMS 4$t$ search on $\rho_{tc}$, and illustrate with our BPs.

\subsection{\boldmath Constraints from CMS $4t$-CRW}\label{constcrw}

The CMS search for SM $4t$ production~\cite{Sirunyan:2019wxt} is 
based on 137~fb$^{-1}$ data at $\sqrt{s}=13$ TeV, i.e. with full Run~2 data.
Based on the number of $b$-tagged jets and charged leptons ($e$, $\mu$), 
the CMS search divides its analysis into 
several signal regions (SRs) and two control regions (CRs). 
The baseline selection criterion requires each event should have 
at least two same-sign leptons.
The remaining selection cuts goes as follows~\cite{Sirunyan:2019wxt}:
The leading and subleading leptons
should have $p_T > 25$ and $20$\;GeV respectively,
with electron (muon) pseudorapidity satisfying
$|\eta| < 2.5$\;($2.4$), whereas all jets should satisfy $|\eta| < 2.4$.
Events are selected if $p_T$ of the jets and $b$-jets fulfill 
any of the following three criteria~\cite{info-Jack}:
(i) both $b$-jets satisfy $p_T$\,$>$\,$40$\;GeV;
(ii) one $b$-jet with $p_T$\,$>$\,$20$\;GeV and $20$\,$<$\,$p_T$\,$<$\,$40$\;GeV
  for the second $b$-jet, but $p_T$\,$>$\,$40$\;GeV for the third jet;
(iii) both $b$-jets should satisfy $20$\,$<$\,$p_T$\,$<$\,$40$\;GeV,
  but with two extra jets each with $p_T$\,$>$\,$40$\;GeV.
Defining $H_T$ as the scalar sum of the $p_T$ of all jets~\cite{Sirunyan:2019wxt}, 
CMS requires $H_T$\,$>$\,$300$\;GeV and $p_T^{\rm miss}$\,$>$\,$ 50$\;GeV.
To reduce the charge-misidentified Drell-Yan ($Z/\gamma^*$) background 
with electrons, events with same-sign electron pairs 
with $m_{ee}$\,$<$\,$12$\;GeV are rejected.
With these selection criteria, the expected total number of events 
(SM backgrounds plus $4t$) in the CRW are $335 \pm 18$, 
with 338 events observed~\cite{Sirunyan:2019wxt}.

It is found~\cite{Hou:2018zmg,Hou:2019gpn,Hou:2020ciy,Hou:2020chc} 
that the most stringent constraint on $\rho_{tc}$ arises from 
the CRW, or $t\bar t W$ control region~\cite{Sirunyan:2019wxt}.
These works studied the $\rho_{tc}$-induced $cg\to tH/tA \to t t \bar c$ processes. 
When both the same-sign top quarks decay semileptonically, 
these processes would contribute to the CRW. 
But the $cg\to b H^+ \to bAW^+$ process with $A \to t \bar c$ 
would also contribute to the CRW if the top decays semileptonically 
and the $W^+$ decays leptonically. 
To estimate the CRW constraints for our BPs, 
one has to add contributions from both $cg\to b H^+ \to bAW^+$ and 
$cg\to tH/tA \to t t \bar c $ coherently, as both effectively give
$cg\to  \ell^+ \ell^+ b b \nu_\ell \nu_\ell \bar c$, which we denote as 
the same-sign dilepton with $2b$ plus extra jet (SS$2l$-$2bj$) signature. 
Due to multiple contributing processes 
that are added at amplitude level, one cannot obtain
simple $\sigma \times \mathcal{B}$ scaling formulas for the BPs. 
Therefore, unlike mass vs $\rho_{tc}$ exclusion contours as in Refs.~\cite{Hou:2018zmg,Hou:2020ciy,Hou:2020chc}, here we test directly 
whether a reference $\rho_{tc}$ value survives the CRW constraint. 
In particular, we take the relatively low $\rho_{tc} =0.15$ for illustration. 
 
Under the aforementioned assumptions on couplings, 
i.e. turning off all other $\rho_{ij}$ except $\rho_{tc}$, 
the total decay width of $H^+$ is the sum of 
$H^+\to c \bar b$ and $H^+ \to A W^+$ partial widths, 
while for $H$ it is the combination of $H\to t \bar c +\bar t c$ 
and $H\to A Z$ decays for both the BPs, 
with $\mathcal{B}(A\to t \bar c +\bar t c)=100\%$. 
For $\rho_{tc} =0.15$, the total decay widths of $A$, $H$ and $H^+$ 
are 2.04 (2.27), 0.029 (0.35) and 2.65 (2.9) GeV for BP$a$ (BP$b$).

We first generate  SS$2l$-$2bj$ events for both the BPs at $\sqrt{s}=13$ TeV 
using MadGraph5\_aMC@NLO~\cite{Alwall:2014hca}
 (denoted as MadGraph5\_aMC) at leading order (LO) 
with default parton distribution function (PDF) set NN23LO1~\cite{Ball:2013hta}, 
and interface with PYTHIA~6.4~\cite{Sjostrand:2006za} for showering and hadronization.
The events are then fed into Delphes~3.4.2~\cite{deFavereau:2013fsa} 
for fast detector simulation. 
Here in our exploratory analysis we use the default CMS-based detector card of 
Delphes~3.4.2 for the CMS CRW to incorporate detector effects such as 
$b$-tagging and light-jet misidentification efficiencies etc.
The jets are reconstructed via anti-$k_T$ algorithm with radius parameter $R=0.5$. 
The effective model is implemented in the FeynRules~\cite{Alloul:2013bka} framework.

Following the same event selection cuts of the CRW, 
the SS$2l$-$2bj$ cross section for the two BPs are 0.283 and 0.245~fb. 
Multiplying by the 137~fb$^{-1}$ integrated luminosity, 
these translate to $\sim 39$ and 34 events, respectively, which
should have shown up already in the CRW of CMS $4t$ search~\cite{Sirunyan:2019wxt}. 
Demanding that the combination of the number of events expected 
from the SM and the $\rho_{tc}$-induced 
same-sign dilepton with $2b$ plus extra jet events 
agree with the observed number of events 
within $2\sigma$ uncertainty of the expected, 
we see that $\rho_{tc} =0.15$ is barely allowed for either BPs. 
We note that $\rho_{tc} \gtrsim 0.15$ is the ballpark exclusion
limit found in  Ref.~\cite{Hou:2020chc} from SS$2l$-$2bj$ 
arising from $cg\to tH/tA \to t t \bar c $ processes alone, 
with a mass hierarchy $m_{H}\sim m_A\sim m_{H^+}$, 
but the $cg\to b H^+ \to bAW^+$ process was not induced. 
This illustrates that SS$2l$-$2bj$ events arising from $cg\to b H^+ \to bAW^+$ is significant, 
and CRW constrains $\rho_{tc}$ more stringently if $m_{H^+} > m_A + m_{W^+}$.
Here we remark that the constraint from CRW is extracted with default CMS based detector
card of Delphes. In our exploratory analysis, we have not validated the results of Ref.~\cite{Sirunyan:2019wxt}
which we leave out for future.
In any case, we would see shortly that a dedicated SS$2l$-$2bj$ search could be more sensitive
than the constraint from CRW.

ATLAS has also made similar search~\cite{Aad:2020klt} 
but due to difference in defining SRs and selection criteria, 
the constraints~\cite{Hou:2020ciy} are found to be weaker than CMS. 
Other searches such as for squark pair production~\cite{Aad:2019ftg},
and for new phenomena with same-sign dileptons and $b$-jets~\cite{Aaboud:2018xpj},
both by ATLAS, have 
too strong selection cuts to give meaningful constraint.

\subsection{A dedicated SS$2l$-$2bj$ search}

Even though the existing CMS $4t$ search with full Run~2 data
can set meaningful constraints on the parameter space,
it is not optimized for $cg\to b H^+ \to bAW^+$  search.
This motivates us to perform a dedicated search for SS$2l$-$2bj$ for our BPs at 14 TeV LHC.
Here, we closely follow the analysis of Ref.~\cite{Hou:2020chc}.

There are several SM backgrounds for a dedicated SS$2l$-$2bj$ search. 
The dominant ones are $t\bar t Z$, $t\bar t W$, 
with $4t$, $t\bar t h$ and $tZ +$\,jets subdominant.
In addition,  if the lepton charge gets misidentified 
(charge- or $Q$-flip), with misidentification efficiency at
$2.2\times 10^{-4}$~\cite{ATLAS:2016kjm,Aaboud:2018xpj,Alvarez:2016nrz},
the $t\bar t+$\,jets and $Z/\gamma^*+$\,jets processes would also contribute.
We remark that the CMS study~\cite{Sirunyan:2017uyt} 
with similar final state topology but with slightly different cuts 
finds ``nonprompt'' backgrounds at $\sim 1.5$ times 
the $t\bar{t}W$ background, which is significant.
As the nonprompt backgrounds are not properly modeled in Monte Carlo simulations, 
we simply add this component to the overall background 
at 1.5 times the $t\bar t W$ background after selection cuts.
There are also some tiny backgrounds such as $3t + W$ and $3t + j$, 
which we neglect in our analysis. 

For generating signal and background event samples, 
we follow the procedure as in the previous section and adopt 
MLM matching~\cite{Mangano:2006rw,Alwall:2007fs} prescription 
for matrix element and parton shower merging. 
We allow one additional parton for $t\bar t Z$, $t\bar t W$ and $t\bar t+$\,jets, 
while for other backgrounds and the signal we do not consider additional partons. 
This restriction is due to computational limitations in this first attempt,
and we adopt default ATLAS based detector card of Delphes 3.4.2.

The LO cross sections of $t\bar t Z$, $t\bar{t} W^-$\,($t\bar{t} W^+$),  
$4t$, $t\bar t h$ and $tZ+$\,jets are normalized to next-to-leading order (NLO) 
by the factors 
1.56~\cite{Campbell:2013yla}, 1.35 (1.27)~\cite{Campbell:2012dh}, 
2.04~\cite{Alwall:2014hca}, 1.27~\cite{twikittbarh} and 
1.44~\cite{Alwall:2014hca}, respectively.
The same QCD correction factor is taken for 
the charge conjugate $\bar{t}Z+$\,jets background for simplicity.
The $Q$-flip $t\bar t +$\,jets and $Z/\gamma^*+$\,jets components 
are adjusted to NNLO cross sections by factors of 
$1.84$~\cite{twikittbar} and $1.27$, respectively, 
where we use FEWZ 3.1~\cite{Li:2012wna} to obtain the latter.

To reduce backgrounds, we follow a cut based analysis that 
differs from the CRW of Ref.~\cite{Sirunyan:2019wxt}.
The leading and subleading leptons are required to have $p_T > 25$ and 20\,GeV,
respectively, while $|\eta| < 2.5$ for both leptons. 
All three jets should have $p_T> 20$\;GeV, whereas $|\eta| < 2.5$. 
The $E^{\rm miss}_{T}$ in each event should be $> 35$ GeV.
The $\Delta R$ separation between any lepton and any jets ($\Delta R_{\ell j}$), 
between the two $b$-jets ($\Delta R_{bb}$), 
and between the same-sign leptons ($\Delta R_{\ell\ell}$),
should all satisfy $\Delta R > 0.4$. 
Finally, all selected events should have $H_T > 300$\;GeV,
with $H_T$ defined according to ATLAS, 
i.e. including the $p_T$ of the two leading sames-sign leptons. 

The background cross sections after selection cuts
 are summarized in Table~\ref{backg}, while the 
signal cross sections along with significance for 
the corresponding BPs are given in Table~\ref{signi}. 
The significance is computed using the likelihood for 
a simple counting experiment~\cite{Cowan:2010js},
\begin{align}
  Z(n|n_\text{pred})= \sqrt{-2\ln\frac{L(n|n_\text{pred})}{L(n|n)}}, 
 \label{poisso}
\end{align}
with $L(n|\bar{n}) = {e^{-\bar{n}} \bar{n}^n}/{n !}$,
where $n$ ($n_\text{pred}$) is the observed (predicted) number of events. 
For discovery, one compares the signal plus background ($s+b$) 
with the background prediction ($b$) and demand $Z(s+b|b) > 5$, 
while for exclusion we demand $Z(b|s+b) > 2$.

\begin{table}[t!]
\centering
\begin{tabular}{c |l c c c }
\hline

                     \,Backgrounds\,                 & \ \,Cross section (fb)\,     
& \\
\hline
\hline
                       $t\bar{t}W$               & \hskip1.1cm 1.31               \\
                       $t\bar{t}Z$                & \hskip1.1cm 1.97               \\
                       $4t$                           & \hskip1.1cm 0.316               \\
                       $tZ+$\,jets                 & \hskip1.1cm  0.255                \\
                      $t\bar t h$                   & \hskip1.1cm  0.07                \\                       
                     $Q$-flip                       & \hskip1.1cm  0.024                 \\
                     nonprompt                      & \hskip0.8cm $1.5\times t\bar{t}W$      \\
\hline
\hline
\end{tabular}
\caption{
Background cross sections after selection cuts for the dedicated SS$2l$-$2bj$ search.
}
\label{backg}
\end{table}

\begin{table}[t!]
\centering
\begin{tabular}{c |c| l c | c }
\hline 
\hspace{.1cm}  BP   \hspace{.1cm}  \hspace{.1cm} & \hspace{.05cm} \ Signal \  \hspace{.1cm}& \hspace{.06cm} \ Significance ($\mathcal{Z}$) \hspace{.1cm}   \\ 
                                  &   \, (fb)                                &  \quad 300 (3000) fb$^{-1}$    \\ 
\hline
\hline
                     $a$                       & \,0.468                 & \quad\quad\ \;3.3  (10.4)    \\ 
                     $b$                       & \,0.334                 & \quad\quad\ \;2.4 \ \,(7.5)    \\
\hline
\hline
\end{tabular}
\caption{
Signal cross sections and significances with 300 (3000) fb$^{-1}$ 
for the BPs of SS$2l$-$2bj$ search after selection cuts.
}
\label{signi}
\end{table}

We see from Table~\ref{signi} that, for BP$a$ one can reach 
the significance \ of $\sim 3.3\sigma\,(10.4\sigma)$ 
with 300\;(3000) fb$^{-1}$, while correspondingly
$\sim 2.4\sigma~(7.5\sigma)$ for BP$b$. 
Reanalyzing for a reference value of $\rho_{tc}=0.1$, 
we find that significances at $\sim2.8\sigma\,( 4.8\sigma)$, 
$\sim2\sigma\,( 3.5\sigma)$ are possible for BP$a$, BP$b$ 
with 1000 (3000) fb$^{-1}$.
For exclusion,
we find that $\rho_{tc}=0.1$ can be excluded 
for BP$a$ (BP$b$) with 600 (1000) fb$^{-1}$ data. 
Thanks to the presence of the $cg \to bH^+ \to bAW^+$ process,
these are \textit{well below} the exclusion reach of HL-LHC data 
based on the $cg\to tA/tH\to t t\bar c$ process alone, 
as was found in Ref.~\cite{Hou:2020chc}.

\section{Impact on the CMS excess}\label{cmsexces}

In studying the prospect of 
$cg\to b H^+ \to bAW^+ \to bt\bar c W^+ \to W^+W^+bb\bar c$ at the LHC,
because of interference with the $cg \to tH/A \to tt\bar c$ in the same final state,
we find elevated impact. 
Given the correlation~\cite{Hou:2019gpn} of the $cg \to tH/A \to tt\bar c$ process
and the $gg\to A \to t \bar t$ excess hinted by CMS~\cite{Sirunyan:2019wph}, 
the $cg\to b H^+ \to bAW^+$ process should therefore make strong impact 
on the g2HDM interpretation, which we now turn to elucidate.

CMS has reported~\cite{Sirunyan:2019wph} a hint of excess 
in $gg \to H/A \to t \bar t$ resonance search with 35.9 fb$^{-1}$ data at 13 TeV. 
The search fits for a peak and dip structure~\cite{Carena:2016npr} 
in the $t\bar t$ invariant mass ($m_{t\bar t}$) from interference between 
$gg \to H/A \to t \bar t$ and the rather large $gg \to t\bar t$ QCD background.  
A signal-like deviation is reported~\cite{Sirunyan:2019wph} around $m_A =  400$\;GeV 
and $\Gamma_A/m_A = 4\%$  from a model-independent analysis. 
The local significance is $(3.5\pm 0.3) \sigma$, becoming $1.9\sigma$ 
if one takes into account look-elsewhere effect. 
The deviation depends mildly on $\Gamma_A/m_A$, 
while no deviation is seen for the $CP$-even scalar boson $H$.
CMS does not provide the $At\bar t$ coupling strength, 
using instead a ``coupling modifier'' $g_{At\bar t}$~\cite{Sirunyan:2019wph},
which is nothing but $g_{At\bar t} = \rho_{tt}/\lambda_t$ in g2HDM,
%
and one can utilize the supplementary material of 
Ref.~\cite{Sirunyan:2019wph} to infer its value.
Note that ATLAS~\cite{Aaboud:2017hnm} has performed 
a similar search for distorted Breit-Wigner shape in $m_{t\bar t}$ 
with 8 TeV data for $m_{A,H} > 500$\;GeV, with no excess seen.

\begin{table*}[t]
\centering
\begin{tabular}{c |c| c| c| c | c | c| c | c |c| c| c| c}
\hline
$\rho_{tt}$&  $\rho_{tc}$& $\eta_1$ &  $\eta_2$   &  $\eta_3$   & $\eta_4$  & $\eta_5$ & $\eta_6$ 
& $\eta_7$  & $m_{H^+}$  & $m_A$ & $m_H$ &  $\frac{\mu_{22}^2}{v^2}$\\
 &&&&&&&&& (GeV) & (GeV) & (GeV)&\\ 
\hline
  1.1 & 1 &  0.258 & 1.894  & 8.872 & $-4.772$ & 4.752 & 0
 & $-0.514$ & 670 & 400 & 670 & 2.96                         \\
\hline
\hline
\end{tabular}
\caption{Parameter values to interpret the $gg\to A \to t \bar t$ excess hinted by CMS~\cite{Sirunyan:2019wph}.}
\label{param}
\end{table*}

The CMS ``excess'' is rather close to the $t\bar t$ threshold, and one needs 
a better understanding of the interference with signal near threshold,
and even $gg \to t\bar t$ production in SM as well. 
Nevertheless, it is of interest to see whether the ``excess'' 
can be interpreted in g2HDM. 
Taking $g_{At\bar t} = 1.1$ (hence $\rho_{tt}\approx1.1$), 
Ref.~\cite{Hou:2019gpn} treated the 95\% C.L. upper limit 
at $m_A = 400$\,GeV with $\Gamma_A/m_A = 5\%$ 
as the closest (among the six plots given in Ref.~\cite{Sirunyan:2019wph}) 
to the reported 3.5$\sigma$ excess with $\Gamma_A/m_A = 4\%$, 
but it would have been preferable to have CMS provide 
the coupling modifier value for the excess.
It was found~\cite{Hou:2019gpn} that the excess at $m_A = 400$\,GeV 
with $\rho_{tt}\sim 1.1$ can be compatible with $\rho_{tc}\sim0.9$ and 
$m_H\gtrsim 500$\,GeV, $m_{H^+}\gtrsim 530$\,GeV in g2HDM.
%
This took into account various constraints similar to those considered
in Sec.~\ref{sec:Constraints},
plus
$pp \to \bar t (b) H^+ \to \bar t (b) t \bar b$ searches, 
and 
also neutral Higgs searches such as 
$pp \to t \bar t A/t \bar t H \to t\bar t t \bar t$~\cite{Sirunyan:2019wxt}, as well as
the limits from $gg\to H\to t \bar t$ searches by 
CMS~\cite{Sirunyan:2019wph} and ATLAS~\cite{Aaboud:2017hnm}.
%
These need to be retraced with adding the amplitude induced by $cg\to b H^+ \to bAW^+$.

But before that, we remark that the sizable $\rho_{tc}\sim0.9$ value
plays a mutually compensating role with the large $\rho_{tt} \sim 1.1$ 
needed to account for the CMS excess.
Sizable $\rho_{tc}$ dilutes ${\cal B}(A/H\to t \bar t)$ (${\cal B}(H^+ \to t \bar b$)) 
by $A/H \to t \bar c + \bar t c$ ($H^+\to c \bar b$) decays,
hence weakens the constraints from $pp \to t \bar t A/t \bar t H \to t\bar t t \bar t$ 
and $gg\to H\to t \bar t$ ($pp \to \bar t (b) H^+ \to \bar t (b) t \bar b$) searches. 
In turn, $\rho_{tt}\sim 1.1$ helps alleviate the constraint on $\rho_{tc}$ 
from $cg\to tA/tH \to t t \bar c$ by finite $\mathcal{B}(A/H\to t \bar t)$. 
The most stringent constraint arises from SR12 of CMS~\cite{Sirunyan:2019wxt} search, 
the signal region (SR) for SM $4t$ production, 
defined as at least three charged leptons ($e$, $\mu$), 
three $b$-tagged jets but restricting to four jets, plus some $E_T^{\rm{miss}}$. 
CMS observed 2 events in SR12 in the cut-based analysis 
whereas $2.62 \pm 0.54$ events were expected~\cite{Sirunyan:2019wxt}.
With both $\rho_{tt}\sim 1.1$ and $\rho_{tc}\sim0.9$, 
the $cg\to tA/tH\to t t \bar t$ process would contribute to SR12 
if all three top decays semileptonically, 
but it was found to be compatible with SR12~\cite{Hou:2019gpn}.

Most constraints analyzed in Ref.~\cite{Hou:2019gpn} remain the same, 
but new LHC  results on $pp \to \bar t (b) H^+ \to \bar t (b) t \bar b$ search
became available~\cite{ATLAS:2020jqj,Sirunyan:2020hwv} 
and seem to push $m_{H^+}$ toward the heavier side, making the benchmark
point analyzed in Ref.~\cite{Hou:2019gpn} incompatible with the excess. 
We find a new allowed benchmark point, summarized in Table~\ref{param}, 
that can account for the excess while satisfying perturbativity, unitarity, positivity 
and electroweak precision measurements (checked via 2HDMC~\cite{Eriksson:2009ws}),
as well as all experimental constraints described in Ref.~\cite{Hou:2019gpn}, 
while taking into account the new results from 
Refs.~\cite{Sirunyan:2020hwv,ATLAS:2020jqj}.  
The total widths of $A$, $H$ and $H^+$ for this BP 
are $\sim 30$, 87 and 105\,GeV, respectively.
The respective branching ratios are 
$A\to t \bar t$ and $t \bar c + \bar t c \approx 48\%$ and $52\%$; 
$H\to t \bar t,\;  t \bar c + \bar t c$ and $A Z
 \approx 35\%$, $40\%$ and $25\%$; and
$H^+\to  t \bar b,\; c \bar b$ and $A W^+ \approx 40\%$, $38\%$ and $22\%$.
We neglect tiny loop induced decays for simplicity.

The spectrum in Table~\ref{param} would again allow 
the $cg\to b H^+ \to bAW^+$ process, and therefore contribute to 
the CRW of Ref.~\cite{Sirunyan:2019wxt}. 
We generate SS$2l$-$2bj$ events 
from $cg\to b H^+ \to bAW^+$ and $cg \to t A/tH \to t t \bar c$ 
 at $\sqrt{s}=13$\,TeV for this BP. 
Following the same selection criteria and procedure as described in Sec.~\ref{constcrw}, 
we find the SS$2l$-$2bj$ cross section to be $\sim 1.3$\;fb. 
Multiplying by 137 fb$^{-1}$ integrated luminosity, 
this translates to an overwhelming 179 events. 
This suggests that the BP and hence the g2HDM interpretation of the CMS excess 
is already in severe tension with the CRW of Ref.~\cite{Sirunyan:2019wxt}.
At this point we also remark that the BP in Table~\ref{param} has twisted custodial symmetry 
i.e. $m_{H^+}=m_H$, which helped us evade stringent electroweak precision observables such as $T$ parameter. 
In general mass splitting between $m_H$ and $m_{H^+}$ is possible however, such choice would lead to stringent
constraints from electroweak precision observables. In addition, for lighter $m_H$, 
in particular for $m_{H^+} > m_H + m_{W^+}$ the $cg\to b H^+ \to b H W^+$ process with $H\to t \bar c$ decay
would also induce SS$2l$-$2bj$ signature and contribute to CRW region.

Presence of other $\rho_{ij}$'s may reduce the required $\rho_{tc}$ for the excess, 
but would be subject to other stringent constraints from flavor physics and LHC. 
For example, $\rho_{tu}$ can still be sizable~\cite{Hou:2020ciy}, 
which would also induce SS$2l$-$2bj$ events 
via $V_{tb}$-enhanced $ug\to b H^+ \to bAW^+$ process, 
as well as the $ug \to t A/tH \to t t \bar u$ process. 
Having both $\rho_{tu}$ and $\rho_{tc}$, one would need to consider stringent 
constraints from $D$--$\overline{D}$ mixing~\cite{Crivellin:2013wna,Altunkaynak:2015twa}. 
A nonvanishing $\rho_{\tau\tau}$ may help reduce the requirement of large $\rho_{tc}$.
However, together with $\rho_{tt}$, such parameter space would also receive 
several meaningful constraints from flavor physics and low energy observables
 (see e.g Refs.~\cite{Crivellin:2013wna,Omura:2015xcg,Iguro:2017ysu,Hou:2020itz}).
Presence of $\rho_{bb}$ would make the situation worse via $V_{tb}$ enhanced 
$pp \to \bar t (b) H^+ \to \bar t (b) t \bar b$ process, in addition to other stringent 
constraints as discussed in Refs.~\cite{Modak:2019nzl,Modak:2020uyq}.
We therefore do not think in its minimal set up g2HDM can explain the
CMS hint for an excess at $m_A \approx 400$ GeV.

\section{Discussion and Outlook}\label{disc}

We have analyzed the possibility of probing the FCNH coupling $\rho_{tc}$ 
at the LHC via the $cg \to b H^+ \to bAW^+$ process at 14 TeV LHC.
With the novel signature of same-sign dilepton plus $2b$ and an extra jet  (SS$2l$-$2bj$), 
the process can be discovered even for $\rho_{tc}$ down to 0.1. 

Some uncertainties in our results have not been covered.
The $c$-quark initiated $cg \to bH^+$, $t A/t H$ processes 
have non-negligible  systematic uncertainties such as from PDF and scale dependence 
 (see e.g. Refs.~\cite{Buza:1996wv,Maltoni:2012pa,Butterworth:2015oua}),
which we did not include in our analysis. 
Moreover, we have not included nonprompt and fake backgrounds. 
These induce some uncertainties in our results.

The FCNH coupling $\rho_{tu}$, as mentioned, 
can also induce similar final state topologies via the $ug \to b H^+\to bAW^+$ process. 
One may distinguish between the $\rho_{tc}$ 
and $\rho_{tu}$ induced processes via charge asymmetry of positively
and negatively charged dilepton signature, as discussed in~\cite{Hou:2020ciy}.
Presence of both $\rho_{tc}$ and $\rho_{tu}$ can obscure the role of each other. 
However, in such a case, $D$--$\overline{D}$ mixing can 
provide some probe~\cite{Crivellin:2013wna,Altunkaynak:2015twa}.
For example, Ref.~\cite{Altunkaynak:2015twa} found that 
$|\rho_{tu}^*\rho_{tc}| \gtrsim 0.02$ could be excluded by $D$--$\overline{D}$ mixing 
for $m_H \approx m_A \approx m_{H^+} \simeq 500$\;GeV. 
Moreover, nonzero $\rho_{tu}$, with the help of nonzero $\rho_{\tau\mu}$, 
can induce observable effects in the branching ratio of $B\to \mu \nu$~\cite{Hou:2019uxa},
which is within reach of Belle-II~\cite{Kou:2018nap}.

We have focused mainly on the parameter space where 
all other $\rho_{ij}$'s vanish. 
However, the $\rho_{ij}$ couplings would likely share~\cite{Hou:1991un,Hou:2017hiw} 
the same flavor organization as in SM, 
i.e. $\rho_{ii}\sim \lambda_i$, while off-diagonal elements trickle off. 
This would suppress the discovery potential of SS$2l$-$2bj$ signature to some extent,
where we have discussed the impact of $\mathcal{O}(1)$ $\rho_{tt}$ in Sec.~\ref{cmsexces}.
Finite $\rho_{tt}$ actually motivates conventional searches 
such as $gg\to H,A\to t \bar t$ and $gg\to H t \bar t\to t \bar t t \bar t$~\cite{4top}.
Furthermore, if $\rho_{tc}$ and $\rho_{tt}$ are both finite, 
one may have the more exquisite $cg\to tA/tH \to t t \bar t$~\cite{Kohda:2017fkn}
and $cg\to b H^+\to b t \bar b$ processes~\cite{Ghosh:2019exx},
where the latter may emerge in LHC Run~3.

We find that the SS$2l$-$2bj$ signature arising from 
$cg \to b H^+ \to bAW^+$ and $cg\to tA/tH \to t t \bar c$ processes 
together can exclude a g2HDM interpretation of the $gg \to A \to t \bar t$ 
``excess'' hinted by CMS~\cite{Sirunyan:2019wxt}.
One may push $H^+$ to avoid such constraint, but this should also be
tightly constrained by electroweak precision measurements as well as perturbativity. 
The latter tension can be readily seen from the $\eta_3$ value in Table~\ref{param}. 
Presence of multiple nonvanishing $\rho_{ij}$ may help alleviate the tension.
However, we remark that such an effort would require 
a more involved analysis, which we leave for the future.

In summary, we have analyzed the prospect for discovering 
the $cg \to b H^+ \to bAW^+$ process at the 14\;TeV LHC,
and show that it receives stringent constraint from 
some control region of the existing CMS $4t$ search.
We find that a dedicate search with the signature of same-sign dilepton, 
two $b$-tagged jets plus an additional jet and missing transverse energy 
can provide better probe of the parameter space.
The process can essentially exclude the g2HDM explanation of 
$gg\to A \to t \bar t$ excess observed by CMS. 
If the $cg \to b H^+ \to bAW^+$ process is discovered,
it would not only confirm the existence of new physics, 
it may also help us understand the mechanism behind 
the observed baryon asymmetry of the Universe.

\vskip0.2cm
\noindent{\bf Acknowledgments.--} \
The work of TM is supported by a Postdoctoral Research Fellowship from 
the Alexander von Humboldt Foundation.
The work of WSH is supported by MOST 109-2112-M-002-015-MY3 of Taiwan
and NTU 110L104019 and 110L892101. 


\end{document}